\documentclass[amsmath,superscriptaddress,amssymb]{revtex4}
\begin{document}

\def\beqa{\begin{eqnarray}}
\def\eeqa{\end{eqnarray}}
\def\beqn{\begin{equation}}
\def\eeqn{\end{equation}}

\def\g{g}
\def\wz{w}
\def\D{D}                 

\def\s{s}                  
\def\sx{\s_1}
\def\sxm{\s_1^-}
\def\sxp{\s_1^+}
\def\sy{\s_2}      
\def\x{x}                  
\def\t{t}                  
\def\tx{\t_1}
\def\txm{\t_1^-}
\def\txp{\t_1^+}
\def\ty{\t_2}
\def\tz{\tau}
\def\Dt{\Delta \t}
\def\xb{\mathbf{x}}       
\def\xbx{\xb_1}
\def\xbxm{{\xb_1^-}}
\def\xbxp{{\xb_1^+}}
\def\xby{{\xb_2}}
\def\r{r}                  
\def\rx{{\r_1}}
\def\ry{{\r_2}}
\def\ryz{{\r_0}}
\def\ang{\varphi}          
\def\angx{{\ang_1}}
\def\angy{{\ang_2}}
\def\angxy{\phi} 
\def\angl{\theta}
\def\anglx{\angl_1}
\def\angly{\angl_2}
\def\angt{{\tilde \ang}}
\def\anglt{{\tilde \angl}}

\def\u{u}               
\def\U{U}               
\def\ux{\u_1}  
\def\uxm{\u_1^-}
\def\uxp{\u_1^+}
\def\uy{\u_2}  
\def\ryd{\dot{\r}_2}
\def\rydd{\ddot{\r}_2}
\def\dang{\dot{\angxy}}
\def\dangx{{\dot\angx}}
\def\dDt{\dot{\Delta}\t}
\def\ddDt{\ddot{\Delta}\t}
\def\dty{\partial_\t{\t}_2}
\def\angxyd{{\partial_2\angxy}}
\def\angxyp{{\partial_1\angxy}}

\def\db{{\bar \delta}}

\def\bE{\beta}
\def\cE{\gamma}

\def\c{c}                  
\def\M{M}                  
\def\m{m}                  
\def\jr{j}                 
\def\kr{k}                 
\def\er{e}                 
\def\enr{\varepsilon}      
\def\vlim{\upsilon}        
\def\Gauss{\kappa}
\def\GN{G_N}
\def\Ox{\Omega_1}

\def\chP{\chi_P}            
\def\aP{a_P}                
\def\aE{a_1}                
\def\lP{\ell_P}             
\def\mP{\mu_P}              
\def\rP{r_P}                
\def\PN{\Phi_N}             
\def\PP{\Phi_P}             

\def\td{{\cal T}}               
\def\Td{T}
\def\tdan{\delta \td}
\def\tdr{{\cal C}}              
\def\tdp{{\td_\ang}}               
\def\tdpp{{\td_{\ang\ang}}}
\def\tdp{{\partial_1\td}}
\def\tdd{\partial_2\td}
\def\tddp{\partial_1\partial_2\td}
\def\tddd{\partial_2^2\td}
\def\ri{\rho}                    
\def\rid{\dot{\ri}}
\def\rd{\dot{\r}}
\def\rip{\partial_\angxy\ri}

\def\ynu{y}
\def\ynuan{\delta \ynu}
\def\vd{V}                  
\def\ad{A}                  
\def\annfactor{\mathcal{A}}
\def\tin{_{*}}
\def\tconj{_\mathrm{conj}}

\def\stand#1{\left[#1\right]_\mathrm{st}}
\def\ln{\mathrm{ln}}       
\def\sect#1{sect.\,#1}       
\def\model{{\rm{model}}}
\def\sec{{\rm sec}}
\def\amb{{\rm amb}}
\def\ann{{\rm ann}}

\title{Radar ranging and Doppler tracking in post-Einsteinian metric theories of gravity}
\author{Marc-Thierry Jaekel}
\address{Laboratoire de Physique Th\'eorique de l'Ecole Normale Sup\'{e}rieure, 
CNRS, UPMC, 24 rue Lhomond, F75231 Paris Cedex 05}
\author{Serge Reynaud}
\address{ Laboratoire Kastler Brossel, Universit\'{e} Pierre et
Marie Curie, case 74, CNRS, ENS, Campus Jussieu, F75252 Paris Cedex 05}


\begin{abstract}
The study of post-Einsteinian metric extensions of general relativity (GR),
which preserve the metric interpretation of gravity while considering 
metrics which may differ from that predicted by GR, is pushed one step further.
We give a complete description of radar ranging and Doppler tracking in terms of the 
time delay affecting an electromagnetic signal travelling between the Earth and a remote probe.
Results of previous publications concerning the Pioneer anomaly are corrected 
and an annually modulated anomaly is predicted besides the secular anomaly.
Their correlation is shown to play an important role  when extracting 
reliable information from Pioneer observations.
The formalism developed here provides a basis for a quantitative analysis of 
the Pioneer data, in order to assess whether extended metric theories can be 
the appropriate description of gravity in the solar system. 
\end{abstract}
\maketitle

\section{Introduction}

Experimental tests of gravity show a good agreement with General Relativity (GR)
at all scales ranging from laboratory to the size of the solar system
\cite{Will,Fischbach,Adelberger,JR04}. 
However there exist a few anomalies which may be seen as challenging GR.
Anomalies in the rotation curves of galaxies or 
in the relation between redshifts and luminosities can be accounted for
by considering dark matter and dark energy but they can as 
well be thought of as consequences of modifications 
of GR at galactic or cosmological scales \cite{Lue04,Turner04}.

The anomalous acceleration recorded on Pioneer 10/11 
probes might point at some anomalous behaviour of gravity at a scale 
of the order of the size of the solar system \cite{Anderson98,Anderson02}.
The observation of such an effect has stimulated a significant effort
to find explanations in terms of systematic effects on board the spacecraft 
or in its environment but this effort has not met success up to now \cite{Anderson03}.
The Pioneer anomaly remains the subject of intensive investigation because of its
potential implications in deep space navigation as well as fundamental physics 
\cite{Nieto04,Turyshev04,Bertolami04,Lammerzahl06}.
New missions have been proposed \cite{Pioneer05} and efforts have been made
for recovering data associated with the whole duration of Pioneer 10/11 missions 
and submitting them to new analysis \cite{Nieto05,Turyshev06}.

The present paper follows up publications which have investigated whether or not
metric extensions of GR had the capability to account for the Pioneer anomaly 
while remaining compatible with other gravity tests performed in the solar system. 
Such `post-Einsteinian' extensions preserve the very core of GR with gravity 
identified with the metric tensor $\g_{\mu\nu}$ and motions described by geodesics. 
In particular, the weak equivalence principle, one of the most accurately 
verified properties in physics, is preserved.
However the extended metric may differ from its standard (GR) form so that 
observations may show deviations from standard expectations. 
An important point is that these extensions explore a broader family of metrics 
than in the usually considered PPN family \cite{Will}, including in particular 
deviations in the outer solar system.

These extensions have been introduced in the context of a linearized treatment 
of gravitation fields \cite{JR05mpl,JR05cqg} and then discussed
with non linearity taken into account \cite{JR06cqg}. 
We will show that the previous studies were only preliminary and that 
the more precise and detailed investigations presented in this paper 
change some of their conclusions. 
But the main result, namely that the post-Einsteinian extensions of GR 
show the capability to account for the Pioneer anomaly, will not be affected.
Objections to this statement, contained in recent publications \cite{Iorio06,Tangen06},
will be shown to miss their target. 

The theoretical motivations for extensions of GR are rooted in its long 
confrontation with Quantum Field Theory. Their discussion, presented in 
\cite{JR05cqg,JR06cqg}, is not repeated in the present paper.
Here we will focus our attention on the phenomenological implications of these extensions,
by testing the metric in the solar system through its confrontation with observations, 
particularly those associated with Doppler tracking of Pioneer 10/11 probes. 
These Pioneer data show an anomalous acceleration $\aP$ directed towards the Sun 
with a roughly constant amplitude over a large range of heliocentric distances 
\beqa
\label{Pio_anom}
\aP\sim 0.8\,\mathrm{nm\ s}^{-2} 
\quad,\quad 20\,\mathrm{AU}\le\rP\le70\,\mathrm{AU} 
\eeqa
Note that the positive sign for an acceleration directed towards the Sun has been chosen
to fit the convention of \cite{Anderson02}. The numbers are given as indications
which will allow us to discuss orders of magnitude later on. The symbol AU stands for 
the astronomical unit.

Besides this secular term, the recorded anomalous acceleration also shows diurnal and 
annual modulations \cite{Anderson02}. As the secular one, these modulated anomalies 
could be the consequence of some not yet understood artefact. 
But the search for an artefact accounting for the secular anomaly is usually focused 
on systematic effects on board the spacecraft or in its local environment and
it is clear that modulated anomalies can certainly not be 
explained in this manner, since nothing in the vicinity of Pioneer probes 
is expected to have diurnal or annual variations. 
This entails that secular and modulated anomalies can hardly be due to the same
artefact. 

The main result of the present paper will be that modulated anomalies 
are a natural prediction of post-Einsteinian extensions of GR. 
As a matter of fact, the Doppler observable not only depends on the motion of the 
Pioneer probe but also on the perturbation of electromagnetic propagation 
along the up- and down-links. As the paths followed by these links are themselves 
modulated by motions of the stations, the anomalous Doppler 
acceleration is expected to contain diurnal and annual modulations.
The diurnal and annual anomalies have to be considered as further 
observables of great interest to be confronted to theoretical expectations. 
As these observables can be correlated with the more frequently discussed secular 
anomaly, this opens new perspectives for testing the metric in the solar system, 
even if the systematics associated with modulated and secular anomalies are 
likely not correlated to each other.

In the following, we will give a common description of the secular and modulated 
parts of the anomalous acceleration by introducing a representation of the 
Doppler tracking observables in terms of propagation time delays. 
The advantage of this representation will be to treat in a natural and consistent 
manner the influence of metric perturbations on probe motion on one hand, link
propagation on the other hand. 
The benefit will appear clearly in the discussion of Doppler observables,
deduced by differentiation of the so called radar ranges, that is to say time delays 
between emission and reception, as well as in the interpretation of observations.
We have to stress at this point that Pioneer 10/11 missions were not equipped 
with range measurement capabilities, which is quite unfortunate. 
This indeed leads to ambiguities in the determination of ranges and will be shown
to play an important role in the interpretation of the anomalies. 

Basic definitions and relations between the various quantities will be written down
in the context of the `post-Einsteinian' extensions of GR in the next section (\sect{2}).
The delicate problem of taking into account motions of Earth and probe
will then be addressed (\sect{3}). Exact relations will be presented as well as 
analytical approximations accurate enough for the purpose of the present paper.
We will use the fact that the deviation of the extensions from GR certainly 
remains small since most gravity tests are compatible with GR (\sect{4}).
Using these theoretical tools, we will study the Pioneer-like anomalies possibly 
arising in Doppler tracking of probes in the solar system (\sect{5}). 
In order to discuss the relevant orders of magnitude, we will then make simplifying
assumptions, considering the case of probes moving in the ecliptic plane and having nearly radial 
motions in the outer solar system. We will present theoretical expectations for the 
secular anomaly as well as for the modulated anomaly due to the motion of the Earth 
(\sect{6}), taking into account the correlation arising between these two anomalies.
We will then draw conclusions (\sect{7}) from the results of this new analysis.

\section{Basic definitions and relations}

As already discussed, the high accuracy of tests of the weak principle of equivalence 
allows us to focus our attention on metric extensions of GR. This does not mean that 
there are no violations of this principle but only that such violations are
too small to account for the large Pioneer anomaly (of the order of one thousandth
of the Newton acceleration at the place explored by Pioneer probes).

We also disregard the effects of rotation and non sphericity of the Sun which have
an influence in the inner solar system but hardly in the outer one. 
Hence, we consider a static and isotropic metric representing 
space-time around a punctual and stationary source.
This assumption notably simplifies the description with metric fields only 
depending on two functions $\g_{00}$ and $\g_{\r\r}$ of a single variable, 
the radius $\r$, 
\beqa
\label{Eddington_isotropic_metric}
&&d\s^2 = \g_{00}(\r)\c^2 d\t^2 + \g_{\r\r}(\r)\left( d\r^2
+ \r^2 d\theta^2 + \r^2 \sin^2\theta d\varphi^2\right)\nonumber\\
&&\partial_0 \g_{\mu\nu} = \partial_\theta \g_{\mu\nu} =
\partial_\varphi \g_{\mu\nu} = 0 
\eeqa
The metric has been written with Eddington isotropic coordinates recommended 
by the IAU convention \cite{Guinot,Petit} to represent coordinates in the 
solar system \cite{Schw}. Radii are defined from the gravity source 
($\r \equiv 0$ at Sun center), colatitude angles $\angl$ are defined with respect 
to the ecliptic plane ($\angl \equiv \pi/2$), and azimuth angles $\ang$ 
describe rotation within the ecliptic plane.
For simplicity, we consider the Earth center of motion 
to have a uniform circular motion at frequency $\Ox\equiv2\pi$ yr$^{-1}$. 
We also disregard the problems 
associated with diurnal rotation of Earth and atmospheric perturbations.
These simplifications lead to the drawback that diurnal modulations
will not be modeled.

For the sake of precision, the following remarks have to be made with respect to the IAU convention. 
As already stated, we have first disregarded the effects of rotation and non sphericity 
of the Sun, since they have a small influence on the observables studied thereafter. 
The idea is that differences between the values calculated with the standard 
metric of GR or with the modified metric (\ref{Eddington_isotropic_metric}),
are only slightly affected by this simplification.
The preliminary evaluation of anomalies performed in this manner will have to 
be confirmed by more complete calculations taking into account perturbations due to the 
structure and rotation of the Sun as well as to the presence of planets \cite{Anderson02}.

Then, the IAU convention \cite{Guinot} explicitly refers to GR 
whereas we are here considering extensions of GR. 
We then have to face the implications for the definition of 
fundamental constants, primarily the velocity of light $\c$.
To this aim, we write the metric components as sums
of standard GR expressions and small deviations
\begin{equation}
\g_{\mu\nu} \equiv \stand{\g_{\mu\nu}} + \delta\g_{\mu\nu}   \quad ,\quad 
\vert\delta\g_{\mu\nu} \vert \ll 1
\end{equation}
and we convene that the deviations vanish at the radius of Earth orbit 
on which or in the vicinity of which the most accurate experiments
are performed 
\beqn
\label{null_at_Earth}
\delta\g_{\mu\nu} (r_{1})\equiv 0 \quad ,\quad 
r_{1}\equiv 1\,\mathrm{AU}  \sim 150 \,\mathrm{Gm} 
\eeqn
It thus remains to study the effects of variations with the radius $\r$ 
of the anomalous metric components $\delta\g_{00}$ and $\delta\g_{\r\r}$. 
The standard metric is the GR solution written with Eddington
isotropic coordinates
\beqa
\label{metric_st}
&&\stand{\g_{00}} = \left({1-{\Gauss\over2\r}\over1+{\Gauss\over2\r}}\right)^2 
\quad , \quad \stand{\g_{\r\r}} = - \left(1+{\Gauss\over2\r}\right)^4 
\eeqa
The constant $\Gauss$ is related to the Schwartzschild radius
(with $\GN$ the Newton constant)
\beqa
\label{Gauss}
&&\Gauss \equiv {\GN \M \over \c^2} \sim 1.5 \,\mathrm{km}\sim 10^{-8} \,\mathrm{AU} 
\eeqa
The dimensionless potential $\Gauss/\r$ is small in the solar system, 
with a value $\sim 10^{-8}$ on Earth and even smaller values at the large radii 
explored by Pioneer probes. 

The extensions of GR are often discussed within the PPN framework \cite{Will}
where the metric (\ref{metric_st}) is expanded in terms of the Newton potential 
$\Gauss/\r$ and Eddington parameters $\bE$ and $\cE$ inserted in front of 
the terms of the expansion (with $\bE=\cE=1$ in GR) 
\beqa
\label{Eddington_PPN}
&&\g_{00} =1 - 2 {\Gauss\over\r} + 2\bE {\Gauss^2 \over\r^2} +\ldots 
\quad , \quad \g_{\r\r} = -1 - 2\cE {\Gauss \over\r} - \ldots 
\quad , \quad \mathrm{PPN}
\eeqa
The PPN metric can be considered as a particular `post-Einsteinian' extension of GR 
with anomalies showing specific dependences on the radius 
\beqa
\label{delta_Eddington_PPN}
&&\delta\g_{00} \simeq 2(\bE-1) {\Gauss^2 \over\r^2}  
\quad , \quad \delta\left(\g_{00} \g_{\r\r}\right) \simeq - 2 (\cE -1) {\Gauss \over\r}  
\quad , \quad \mathrm{PPN}
\eeqa
In this paper, we consider more general extensions of GR, which show significant deviations at 
long ranges (outer solar system) and not only short ones (inner solar system). 
With respect to the PPN metric (\ref{delta_Eddington_PPN}), the more general extensions 
can be thought of as allowing for anomalies in the two sectors which may depend on scale.

Einstein curvatures corresponding to the extended metric have been studied
in a detailed manner in \cite{JR06cqg}. 
In contrast to the standard expressions which vanish everywhere except on the gravity 
source, the anomalous curvatures are generally non null in space outside the source.
This is already true for PPN extensions and, again, the general case
corresponds to a more general $\r-$dependence.
This dependence can also be described in terms of running coupling constants
which replace the Newton constant while depending on scale.
We do not repeat these calculations \cite{JR06cqg} but recall that 
the natural metric extension of GR involves 
two running coupling constants which correspond to the sectors of 
traceless and traced tensors \cite{Jaekel95}. 

From the point of view of phenomenology, the two sectors
are as well represented by the two functions $\delta\g_{00}(\r)$ and 
$\delta\left(\g_{00} \g_{\r\r}\right)(\r)$. 
The first sector represents an anomaly of the Newton potential which
has to remain small to preserve the good agreement 
between GR and gravity tests performed on planetary orbits \cite{Fischbach,JR04,Talmadge88}.
Meanwhile, the second sector represents an extension of PPN phenomenology with a scale 
dependent parameter $\cE$. 
It opens an additional phenomenological freedom with respect to the mere modification of the 
Newton potential and this freedom opens the possibility to accomodate a Pioneer-like anomaly 
besides other gravity tests \cite{JR05mpl,JR05cqg,JR06cqg}. 

Recent publications force us to be more specific on the relation between the Pioneer 
anomaly and modifications of the Newton potential, \textit{i.e.} anomalies in the first sector
according to the terminology of the preceding paragraph. Interpreting the Pioneer
anomaly in such a manner requires that $\delta\g_{00}$ varies roughly as $\r$
at the large radii explored by Pioneer probes. If this dependence also holds at smaller 
radii \cite{Anderson02}, or if the anomaly follows a simple Yukawa law \cite{JR04}, 
one deduces that it cannot have escaped detection in the more constraining tests 
performed with martian probes \cite{Reasenberg79,Hellings83,Anderson96}.
Brownstein and Moffat have explored the possibility that the linear dependence 
holds at distances explored by Pioneer probes while being cut 
at the orbital radii of Mars \cite{Brownstein06}. 
Iorio and Giudice \cite{Iorio06} as well as Tangen \cite{Tangen06} have in contrast 
argued that the ephemeris of outer planets were accurate enough to discard the 
presence of the required linear dependence in the range of distances explored 
by the Pioneer probes. 
This argument has been contested by the authors of \cite{Brownstein06}
so that the conflict remains to be settled. 

Iorio and Giudice \cite{Iorio06} and Tangen \cite{Tangen06} have pushed their claim 
one step farther by restating their argument as an objection to the very possibility 
of accounting for the Pioneer anomaly in any viable metric theory of gravity. 
Even before entering the detailed developments to be presented in this paper,
we can show that this claim is untenable just because it only considers metric anomalies
in the first sector while disregarding those in the second sector. 
Later on in the paper, we will come back to the discussion of the compatibility
of the metric modifications with observations performed in the solar system, 
often with an accuracy higher than that of Pioneer observations.
This has to be done with care, accounting for the presence of the two sectors
as well as for possible scale dependences.
This question has already been discussed in \cite{JR05cqg,JR06cqg} for the cases of
deflection experiments on electromagnetic sources passing 
behind the Sun \cite{Iess99,Bertotti03,Shapiro04,Lator,Gaia}. 
We will see below that it has a particularly critical character for the 
ranging experiments which involve directly the Shapiro time delay \cite{Shapiro99}.

\section{Radar ranging and Doppler tracking observables}

In the present section, we introduce the time delay function, a two-point function 
the knowledge of which is equivalent to a characterization of the metric.
We deduce from this function the radar ranging observables between the Earth and 
a probe in the solar system. 
We then analyze the case of Doppler tracking observables
which are obtained from ranging ones through a time differentiation.

Starting from the static and isotropic metric (\ref{Eddington_isotropic_metric})
written in terms of Eddington isotropic coordinates, 
we define the time delay \cite{LePoncin04} as the time taken in this coordinate system 
by a light-like signal to propagate from a spatial position $\xbx$ to another one $\xby$
\beqa
&&\xb_a \equiv \r_a (\sin\angl_a\cos\ang_a,\sin\angl_a\sin\ang_a,\cos\angl_a)
\quad,\quad  a=1,2
\eeqa
This defines a two-point function $\td$ which depends on the positions 
$\xbx$ and $\xby$ only through three real variables, which can be chosen as the two radii 
$\rx$ and $\ry$ and the angle $\angxy$ between the two points as seen from the gravity source, 
\beqa
&&\cos\angxy=\cos\anglx\cos\angly+\sin\anglx\sin\angly\cos\left(\angy-\angx\right) 
\eeqa
Stated differently, the time delay $\td$ is a function of the triangle built on the 
emitter, the receiver and the gravity source.

The form of the time delay function for a static isotropic metric was obtained in 
\cite{JR05cqg,JR06cqg} by solving the Hamilton-Jacobi equation for a light ray
\beqa
\label{time_delay}
\c\td(\rx, \ry, \angxy) &\equiv& \int_\rx^\ry {-{\g_{rr}\over \g_{00}}(\r)
d\r \over \sqrt{-{ \g_{rr}\over \g_{00}}(\r) - {\ri^2\over\r^2}}}
\quad,\quad
\angxy = \int_\rx^{\ry} {\ri d \r/\r^2 \over
\sqrt{-{\g_{rr}\over\g_{00}}(\r) - {\ri^2\over\r^2}}}
\eeqa
These quantities are integrals over the light ray of integrands depending only
on the conformally invariant ratio $\g_{00}/\g_{\r\r}$.
The parameter $\ri$, hereafter called the impact parameter, is an implicit function 
of the variables $\rx, \ry, \angxy$ determined by the second equation in (\ref{time_delay}).
This definition fixes the relative sign of $\ri$ and $\angxy$.

One can also deduce from the time delay function 
\beqa
\label{time_delay2}
&&\c\td = \ri \angxy + \int_\rx^\ry \tdr(\r) d\r \quad, \quad
\tdr(\r) \equiv \sqrt{-{ \g_{rr}\over \g_{00}}(\r) - {\ri^2\over\r^2}}
\eeqa
Calculating the partial derivatives of $\td$, one indeed notices that 
cancellations appear in the angular derivative, due to the particular form
(\ref{time_delay}) of the time delay function,
\beqa
\label{first_derivative}
&&\c\partial_\rx \td = -\tdr(\rx) \quad,\quad
\c\partial_\ry \td = \tdr(\ry) \quad,\quad
\c\partial_\angxy \td = \ri
\eeqa
Second order derivatives of the time delay function $\td$ may then be written
\beqa
\label{second_angular_derivatives}
&&\c\partial^2_\angxy\td = \rip = 
\left(\int_\rx^{\ry} {-{\g_{rr}\over\g_{00}}(\r) d\r/\r^2
\over\tdr(r)^{3}}\right)^{-1} \\
&&\c\partial_\angxy\partial_\ry\td = -{\ri\rip\over\ry^2\tdr(\ry)}= \partial_\ry\ri\quad,\quad
\c\partial^2_\ry \td = \partial_\r\tdr(\ry) + {\ri^2\rip\over\ry^4\tdr(\ry)^2} \nonumber
\eeqa
Similar relations hold involving derivatives with respect to $\rx$ instead of $\ry$.
Note that non local expressions only come from angular derivatives, namely $\ri$ and $\rip$.
It also turns out that the angular second derivative of the time delay function 
is positive (for $\ry>\rx$). 

We now study the observables which are used in deep space navigation \cite{Moyer} 
and make explicit their relation to the time delay function (\ref{time_delay}).
We begin with radar ranging observables obtained by timing radio signals exchanged 
between stations on Earth and the deep space probe.
An up-link radio signal is emitted from Earth at $(\txm, \xbxm)$
and received and sent back by the probe at $(\ty,\xby)$ with the down-link 
radio signal received on Earth at $(\txp,\xbxp)$
(positions identified with the center of motion
of Earth, itself assumed to run uniformly on a circular orbit). 
Meanwhile the deep space probe is assumed to follow a geodesic trajectory
in the outer solar system. As the gravity fields are very small there, this 
trajectory can be approximated with a good approximation as a Keplerian hyperbolic
trajectory escaping the solar system \cite{Anderson02}.

The positions of the emission and reception events are connected by light cones,
that is also by the following relations between event times and the time delay function,
\beqa
\label{lightlike_links}
&&\ty - \txm  =  \td(\xbxm, \xby) \equiv \td_-\quad,\quad
\txp - \ty  =  \td(\xbxp, \xby)  \equiv \td_+
\eeqa
For probes equipped with range measurement capabilities (which was not the case for Pioneer 10/11),
the ranging observable may be defined as half the time elapsed on Earth from the emission time $\txm$
to the reception time $\txp$
\beqa
\label{ranging_time}
\Dt &\equiv& {\txp -\txm \over2} = {\td_- +\td_+ \over2} 
\eeqa
This quantity is not a proper gauge invariant quantity but it is directly related to such a quantity,
the proper time $\sxp-\sxm$ elapsed on Earth between the same two events, through a mere multiplication factor
determined by the potentials created by Sun on Earth and the velocity of Earth on its orbit.
This multiplication factor is constant in the simple context studied in this paper,
and it can therefore be omitted.
The ranging time is available for observers on Earth even if they don't have access to
the transponding time $\ty$ defined on board the deep space probe.
However, the transponding time can be deduced from the solution of equations of motion, at soon as the
metric is known with a sufficient accuracy. We will see below how to deal with this complication.

For Pioneer 10/11 probes, the tracking technique was based on the measurement of the Doppler shift,
a proper observable $\ynu$ defined from the ratio of cycle counting rates 
of reference clocks located at emission and reception stations \cite{Anderson02,Moyer},
\beqa
\label{frequency_shifts}
\ynu &\equiv& \ln{d \sxp \over d\sxm} = \ln{d\txp \over d\txm} 
\eeqa
Note that $\ynu$ has its definition gauge invariant when written in terms of proper times,
but can as well be written in terms of coordinate times on Earth for the same reason as 
in the preceding paragraph.
The same information can be encoded in a Doppler velocity 
\beqa
\label{DopplerVelocity}
&&\frac{\vd}{\c}\equiv \frac{d\Dt}{d\t}=\frac{d\txp-d\txm}{d\txp+d\txm}
=\tanh \frac{\ynu}{2} \quad ,\quad \ynu=\ln \frac{1+\vd/\c}{1-\vd/\c}
\eeqa
The difference between the observables $\vd$ and $\ynu$ appears only at third order in the 
velocities and can be considered as a small term. 
This is due to the choice of the median observer time $t$ to parametrize the data 
\beqa
\label{median_time}
&&\t \equiv {\tx^+ +\tx^-\over2} \quad,\quad
d\txp\equiv d\t+d\Dt \quad,\quad d\txm\equiv d\t-d\Dt 
\eeqa
Should another time be used in its place, we would obtain second order corrections
in the relation between $\vd$ and $\ynu$.

The following remarks are worth being kept in mind when using tracking data for
obtaining knowledge on the motion of the deep space probe.
Clearly the Doppler velocity is primarily correlated to the velocity of the deep space
probe relatively to that of Earth, with relativistic as well as gravitational corrections 
fully accounted for in relation (\ref{DopplerVelocity}). 
But the velocity of the probe at some specific time is not known with a sufficient accuracy,
unless informations extracted from Doppler data are used. 
It follows that the Pioneer gravity test is more appropriately discussed in terms of the 
Doppler acceleration $\ad$, which is just the time derivative of the Doppler velocity 
\cite{Anderson02},
\beqa
\label{DopplerAcceleration}
&&\frac{\ad}{\c}\equiv \frac{d\vd}{\c d\t} = \frac{d^2\Dt}{d\t^2} 
\eeqa
This observable $\ad$ gives a more direct access to the acceleration law of
the probe to be compared with the theoretical expectation.
Now the expected acceleration depends on the distance of the probe to the gravity 
source and the latter also suffers an imprecise determination on probes
for which ranging data are not available. 
These `ambiguity' problems will be faced in the following,
keeping in mind that relativistic and gravitational corrections can in principle
be affected by the discussion.

In order to compare tracking observations with theory, 
we need to make more explicit the relation between the ranging observable $\Dt$
and the time delay function $\td$. To this aim, we note that, as the Earth has been 
assumed to have a circular motion, the time delay function 
$\td$ reduces to a function of two variables $\ry$ and $\angxy$.
For emission and reception events connected by light cones (\ref{lightlike_links}),
these two variables may equivalently be replaced by the time parameters $\tx$ and $\ty$ 
which parametrize the Earth and probe trajectories respectively, at least once
the metric and the trajectories are known.
The history of ranging observations, \textit{i.e.} of series of time triplets  
(see eqs.\ref{lightlike_links}), may then be conveniently represented under the form of a 
relation $\Dt(\t)$ with $\Dt$ and $\t$ defined by (\ref{ranging_time}) and (\ref{median_time}).
The transponding time $\ty$ thus appears as a function of the median time $\t$.
These implicit relations may for instance be solved by approaching in an iterative manner 
the solution of the following equations 
\beqa
\label{ranging_time2}
&&\td_- = \td(\xbx(\t-\Dt),\xby(\ty)) \quad, \quad \td_+ = \td(\xbx(\t+\Dt),\xby(\ty)) \\
&&\ty  = \t + {\td_- - \td_+\over2}\quad,\quad \Dt = {\td_+ + \td_-\over2} \nonumber
\eeqa

Derivatives of the function $\td$ with respect to $\tx$ and $\ty$ are deduced from 
(\ref{first_derivative})
\beqa
\label{time_derivatives}
&&\c\tdp \equiv \c{\partial\td\over \partial\tx} = \ri \angxyp\quad,\quad
\c\tdd \equiv \c{\partial\td\over \partial\ty} = \tdr(\ry)\ryd + \ri \angxyd 
\eeqa
The Doppler velocity and the derivative of transponding time
are given by similar relations
\beqa
\label{synchronization_time_derivative}
&&{c+\vd\over\c-\vd}= {1+\tdd_+\over 1-\tdp_+} {1+\tdp_-\over1-\tdd_-} \quad,\quad 
{2\over\dty} = {1-\tdd_-\over 1+\tdp_-} + {1+\tdd_+ \over 1-\tdp_+} 
\eeqa

\section{Ranging and Doppler anomalies}

In order to bring the relations written in the preceding section to explicit formulas,
we have to face rather complicated expressions, which can be dealt with in a numerical 
procedure but hardly in an analytical calculation.
A simpler approach, extremely useful for a first discussion of the anomalies, 
is to use a first-order expansion of the observables in the metric
perturbations \cite{JR06cqg}.
The basic methods to be used in such an approach are presented in this section.

Let us first discuss the case of the time delay (\ref{time_delay})
with its standard form
\beqa
\label{GR_time_delay}
&&\stand{\c\td}(\rx,\ry,\angxy) = \int_\rx^{\ry} {-\stand{\g_{rr}}\over \stand{\g_{00}}}(\r) 
{d\r \over \stand{\tdr(\r)}}
\eeqa
At first order in the metric perturbation, the post-Einsteinian time delay 
may be written as the sum of this standard form and of an anomaly (see (\ref{time_delay}) and (\ref{time_delay2}))
\beqa
\label{time_delay_anomaly}
&& \c\delta\td (\rx,\ry,\angxy) \equiv \left\lbrace\c\td - \stand{\c\td} \right\rbrace (\rx,\ry,\angxy) 
= \int_\rx^\ry \delta \wz (\r) 
{-\stand{\g_{rr}}\over\stand{\g_{00}}}(\r) {d\r \over \stand{\tdr(\r)}} \nonumber\\
&&\delta \wz \equiv \delta \left\lbrace \ln\sqrt{-\g_{\r\r}\over\g_{00}} \right\rbrace
= {\delta \g_{\r\r}\over 2\stand{\g_{\r\r}}} - {\delta \g_{00}\over 2\stand{\g_{00}}} 
\eeqa
We again notice that this variation is determined by the perturbation of the
conformally invariant ratio $\g_{\r\r}/\g_{00}$. 
In order to compute the variation of the ranging time observable (\ref{lightlike_links}) induced 
by that of the time delay (\ref{time_delay_anomaly}),
we now introduce notations for anomalies of the ranging time $\td$ and transponding time $\ty$
\beqa
&&\delta\Dt \equiv \Dt - \stand{\Dt} \quad,\quad \delta\ty \equiv \ty - \stand{\ty} 
\eeqa
and write a first order equation for these anomalies by linearizing (\ref{ranging_time2})
\beqa
\label{ranging_synchronization_equations}
&&{ \stand{\tdp_+} +\stand{\tdp_-}\over2} \delta\Dt 
+ \left(1+ {\stand{\tdd_+} -\stand{\tdd_-}\over2}\right)\delta\ty
=- {\delta\td_+ - \delta\td_-\over2}\nonumber\\
&&\left(1- {\stand{\tdp_+} -\stand{\tdp_-}\over2}\right) \delta\Dt 
- {\stand{\tdd_+} + \stand{\tdd_-}\over2}\delta\ty
= {\delta\td_+ + \delta\td_-\over2}\nonumber\\
\eeqa
The symbols appearing in (\ref{ranging_synchronization_equations}) are defined according to
\beqa
&&\delta\td_\pm \equiv \delta\td(\xbx(\t \pm \stand{\Dt}),\xby(\stand{\ty})) \\
&&\stand{\partial_a\td_\pm} \equiv \stand{\partial_a\td}(\xbx(\t\pm\stand{\Dt}),\xby(\stand{\ty}))
\quad ,\quad a \equiv 1,2 \nonumber
\eeqa
Solving these equations, one obtains the ranging and transponding time anomalies
\beqa
\label{ranging_synchronization_anomalies}
\delta\Dt &=& {1- \stand{\vd}/\c\over 2}{\delta\td_+\over 1-\stand{\tdp_+}} 
+ {1+ \stand{\vd}/\c\over 2}{\delta\td_-\over 1+\stand{\tdp_-}}\nonumber\\
\delta\ty &=& -{1\over2} \stand{\dty}\left({\delta\td_+\over 1-\stand{\tdp_+}}
-{\delta\td_-\over 1+\stand{\tdp_-}}\right)
\eeqa
The implicit equations (\ref{ranging_time2}) and other expressions written up to now
will be approximated in the next section so that they are more easily used for explicit 
discussions of Pioneer-like anomalies. The approximation will be based on the fact that
the Earth and probe velocities are much smaller than light velocity. 

We now come to the fact that the anomalous time delay between emission and reception is 
not only affected by the perturbation of light propagation, but also by the perturbation 
of the probe trajectory. Precisely, the value of $\ry$ to be used in preceding calculations 
is not the same in extended theory as it would be in standard theory. This is characterized 
by differences $(\delta\ry, \delta\angly, \delta\angy)$ of the coordinates of the probe. 
In order to discuss this point, we introduce a new notation for variations of any quantity $f$
as sums of terms, the first one being calculated with endpoints fixed and the other ones 
associated with position differences,
\beqa
\label{any_variations}
\db f &\equiv& f(\ry, \angxy) - \stand{f}(\stand{\ry}, \stand{\angxy}) 
=\delta f + \delta\ry \partial_\r \stand{f} + \delta\angxy \partial_\angxy \stand{f} 
\eeqa
The angular anomalies have been collected in $\delta\angxy$ and no anomalies have been accounted 
for the position of Earth which is assumed to be known. The condition (\ref{null_at_Earth}) 
of null metric anomalies at Earth ensures the consistency of this description with
the conventions of metrology.
Note that, in a first order expansion in the metric perturbation, all contributions to 
(\ref{any_variations}) except the first one may be calculated using the standard expression of $f$.
For the time delay function in particular, equation (\ref{any_variations}) is read as
\beqa
\label{ranging_anomaly}
\db \Td &\equiv& \delta \td(\stand{\ry}, \stand{\angxy})
+\stand{\tdr}(\stand{\ry}) \delta\ry +\stand{\ri} \delta\angxy
\eeqa
The perturbation $\db \Td $ defined by (\ref{ranging_anomaly}) and evaluated at first order
will be used in the following as a good approximation to the ranging time anomaly 
(\ref{ranging_synchronization_anomalies}).
Its evaluation still requires the solution of the equation of motion of the probe.

In particular, the position differences $\delta\ry$ and $\delta\angxy$ have to be 
deduced from the geodesic equations written separately for the standard metric $\stand{\g}$
and the modified one $\stand{\g}+\delta\g$. 
These geodesic equations have their usual form in a metric theory \cite{Landau}
\beqa
\label{geodesic_equation}
{d\u^\mu \over d\s} + \Gamma^\mu_{\nu\rho} \u^\nu \u^\rho \equiv  0
\eeqa
with $\Gamma^\mu_{\nu\rho}$ the Christoffel symbols 
and $\u^\mu$ the relativistic velocities
\beqa
\label{Christoffel}
\Gamma^\lambda_{\mu\nu} &\equiv& {\g^{\lambda\rho}\over2}
\left(\partial_\mu \g_{\nu\rho} +\partial_\nu \g_{\mu\rho} 
-\partial_\rho \g_{\mu\nu}\right)
\quad,\quad \u^\mu \equiv {d\x^\mu \over d\s }
\quad, \quad \g_{\mu\nu}\u^\mu\u^\nu =1
\eeqa
For the computations to be performed in the next section, it is worth noticing
that the geodesic equations (\ref{geodesic_equation}) may also be written as 
conservation laws for energy and angular momentum. 
The latter is a vector, and the conservation of its direction just means that 
the motion takes place in an orbital plane containing the Sun. 
This plane is characterized by two angles, the longitude of the ascending
node $\Omega $ and the inclination of the orbit $\iota $, which also give 
the spatial direction of the conserved angular momentum along the unit vector
$\left(\sin \iota \sin \Omega , -\sin \iota \cos \Omega , \cos \iota\right)$.

In order to describe motion, we then introduce angular coordinates adapted to 
the orbital plane with $\anglt =\frac{\pi }{2}$ on the orbit, and $\angt$ 
measured in the orbital plane. The reduced energy $\er$ and angular momentum $\jr$
are defined as the following conserved quantities 
\beqa
\label{energy_conservation}
&&\er \equiv \g_{00}{\c d \t\over d\s}\quad,\quad
\jr\equiv \g_{\r\r} \r^2 {d\angt\over d\s}
\eeqa
These relations just give the velocity components $\u^0$ and $\u^\angt$, 
with the third component $\u^\anglt$ vanishing and the fourth one $\u^\r$ 
given by velocity normalization (\ref{Christoffel}) 
\beqa
\label{velocity_normalization}
&&{d\r\over d\s}\equiv\u^\r =\U^\r\quad,\quad
\left(\U^\r\right)^2 \equiv {1\over\g_{00}\g_{\r\r}}\left(\g_{00}
\left(1 -{\jr^2\over\g_{\r\r}\r^2}\right) -\er^2\right) 
\eeqa
$\U^\r$ is a function of the  variable $\r$ giving the radial velocity $\u^\r$ and 
depending on the form of the metric and on the conserved quantities labelling the trajectory.
Equations (\ref{energy_conservation}) can also be written in terms of functions 
$\U^0$ and $\U^\angt$ of the variable $\r$
\beqa
\label{angular_velocities}
&&{cd\t\over d\s}=\U^0\equiv {\er\over\g_{00}} \quad,\quad
{d\angt\over d\s}=\U^\angt\equiv {\jr\over \g_{\r\r} \r^2 }
\eeqa

Standard trajectories are obtained by integrating these equations,
with the metric components having their standard expressions. 
At first order in the metric perturbation around GR, the probe trajectory is then obtained 
as the sum of standard and anomalous contributions. The latter are expressed
in terms of variations of the functions $\U$.
In particular, the radial velocity and acceleration show the following anomalies
\beqa
\label{anomalous_acceleration}
{d\delta \r\over d\s } &=& \delta\left({d\r\over d\s}\right) = \db\U^\r =
\partial_\r\stand{\U^\r} \delta\r + \delta\U^\r 
\nonumber\\
{d^2\delta \r\over d\s^2 } &=& \delta\left({d^2\r\over d\s^2}\right) 
=  \db\left(\U^\r\partial_\r\U^\r\right) =
{1\over2} \partial_\r^2\left\lbrace\stand{\U^\r}^2\right\rbrace  \delta\r +
{1\over2} \partial_\r\delta \left\lbrace\left(\U^\r\right)^2\right\rbrace 
\eeqa
For the explicit calculations to be performed in the next sections, we will
write the solution for the distance variation as
\beqa
\delta\ry&=& \delta\r\tin + \stand{\U^\r(\ry)}  
\int_{\r\tin}^\ry{\delta \U^\r \over\stand{\U^\r}^2} d\r\\
{\delta\U^\r \over\stand{\U^\r }} &=& 
-{\delta(\g_{00}\g_{\r\r}) \over2\stand{\g_{00}\g_{\r\r}}}\nonumber\\
&&+{1\over2\stand{\g_{00}\g_{\r\r} \left(\U^\r\right)^2} }\left\lbrace
{\delta\g_{00}} -{\jr^2\over\r^2} \delta\left({\g_{00}\over\g_{\r\r}}\right)
-{2\jr\delta\jr\over\r^2} {\stand{\g_{00}}\over\stand{\g_{\r\r}}}
-2\er\delta\er \right\rbrace\nonumber
\eeqa
The constant $\delta\r\tin$ represents the initial radius variation at 
 $\r\tin$ between geodesics calculated for the extended and standard metrics.
The angular variation is then written as
\beqa
\delta\angt_2 &=& \delta\angt\tin + 
\stand{\U^\angt(\ry)}\int_{\r\tin}^\ry{\delta \U^\r \over\stand{\U^\r}^2} d\r
+ \int_{\r\tin}^\ry
\left({\delta\U^\angt\over\stand{\U^\angt}} - {\delta\U^\r\over\stand{\U^\r}}\right)
{\stand{\U^\angt}\over \stand{\U^\r}}d\r\nonumber\\
{\delta\U^\angt\over\stand{\U^\angt}}&=& {\delta\jr\over\jr} 
- {\delta\g_{\r\r}\over\stand{\g_{\r\r}}} 
\eeqa
The constant $\delta\angt\tin$ 
represents an initial angular difference between the geodesics.

When taken with the results of the preceding section, these equations provide us with
an exact description of radar ranging and Doppler tracking, in the simplified context 
considered in the present paper and in a first order expansion in the metric deviation from 
its standard form.

\section{Pioneer like anomalies }

Implicit equations written in the preceding section have to be solved in an iterative manner,
which is well adapted to a numerical procedure but not easily performed in an analytical work. 
In order to be able to present qualitative but explicit discussions of Pioneer-like anomalies,
we now introduce approximated forms of these equations.
We will also consider Pioneer-like probes with high excentricity orbits, 
so that it will be possible to neglect angular terms.
For simplicity, we also consider that the probe moves in the ecliptic plane,
\textit{i.e.} that the inclination of the orbital plane vanishes ($\iota=0$).

The main argument pleading for these approximations is the fact that the Earth velocity 
is much smaller than light velocity $\Ox\rx/\c \simeq 10^{-4}$. 
This entails that the change of time delay function due to motion of Earth during
the time of flight of the signal is small.
Furthermore, the parity of equations (\ref{ranging_time2}) and 
(\ref{ranging_synchronization_anomalies}) leads to corrections induced by Earth motion 
appearing only at second order in Earth velocity ($(\Ox\rx/\c)^2 \simeq 10^{-8}$). 
Hence, the modifications of the ranging time anomalies (\ref{ranging_synchronization_anomalies})
due to Earth motion may be ignored in a first discussion.
Note that this statement applies to the anomalous part (\ref{ranging_synchronization_anomalies}) 
of the ranging time and to the effect of motion during the ranging time only.
It does neither hold for the standard ranging time which suffers an appreciable effect
due to the Earth motion \cite{Anderson02}, nor for the effect of Earth motion on a longer term.
The latter effect is evaluated below and found to play a significant role
in the interpretation of the anomaly.

As already discussed, the anomalous time delay is also affected by the perturbation of the 
probe trajectory. This is taken into account by using equation (\ref{ranging_anomaly}) 
which now gives the true anomaly of the ranging observable 
\beqa
\label{true_ranging_anomaly}
\c\db\Dt (\t) &\simeq& \c\db \Td (\t) = \c\delta \td(\rx,  \stand{\ry}, \stand{\angxy})
+\stand{\tdr}(\stand{\ry}) \delta\ry +\stand{\ri} \delta\angxy
\eeqa
The anomaly of Doppler velocity observable is then obtained from (\ref{DopplerVelocity}) and 
(\ref{time_delay_anomaly})
\beqa
\label{Doppler_shift_anomaly}
\db \vd &\simeq& \stand{\ryd}\left\lbrace
{-\stand{\g_{\r\r}}\delta\wz\over\stand{\g_{00}}\stand{\tdr(\ry)}} 
+ \stand{\partial_\r\tdr(\ry)}\delta\ry\right\rbrace
+\stand{\tdr(\ry)}\delta\ryd \nonumber\\
&+& \stand{\ri}\delta\dang + \stand{\rid}{\db\ri\over\stand{\rip}}
\eeqa
where we have introduced shorthand notations
$\ryd \equiv {d\ry\over d\t}$, $\dang \equiv {d\angxy\over d\t}$ and 
$\delta\ryd \equiv {d\delta \ry\over d\t}$, $\delta\dang \equiv {d\delta\angxy\over d\t}$.
The time derivative of the impact parameter appearing in (\ref{Doppler_shift_anomaly}) is given by
\beqa
\rid&=&\left(-{\ryd\ri\over\ry^2\tdr(\ry)}+\dang\right)\rip
\eeqa
All contributions to the Doppler shift anomaly (\ref{Doppler_shift_anomaly}) 
are local except some contributions to the impact parameter anomaly $\db\ri$ 
\beqa
\label{impact_parameter_anomaly}
{\db\ri \over \stand{\rip}} &=& \stand{\ri} \left( \int_\rx^{\ry} 
{-\stand{\g_{\r\r}}\delta \wz \over\stand{\g_{00}}\stand{\tdr}^{3\over2}}{d\r\over\r^2}
- {\delta\ry\over\stand{\ry}^2\stand{\tdr}} \right) +\delta\angxy 
\eeqa
Note that explicit dependences of $\tdr$ on $\ry$ are omitted from now on. 

The anomaly of Doppler acceleration observable is obtained similarly
\beqa
\label{acceleration_anomalies}
\db \ad &=& \stand{\rydd}{-\stand{\g_{\r\r}}\delta\wz\over\stand{\g_{00}}\stand{\tdr}} 
+\stand{\ryd}^2\partial_\r\left({-\stand{\g_{\r\r}}\delta\wz\over\stand{\g_{00}}\stand{\tdr}}\right)\nonumber\\
&&+{d\over d\t}\left(\stand{\ryd}\stand{\partial_\r\tdr}\delta\ry \right)
+{d\stand{\tdr}\over d\t} \delta\ryd +\stand{\tdr} \delta\rydd\nonumber\\
&&+{\stand{\ri}\stand{\dot{\ri}}\stand{\ryd}\over\stand{\ry}^2}
\left({-\stand{\g_{\r\r}}\delta\wz\over\stand{\g_{00}}\stand{\tdr}^3}\right)\nonumber\\
&&+ {d\over d\t}\left\lbrace\stand{\ri}\delta\dang +{\stand{\rid} \over \stand{\rip}} \db\ri\right\rbrace
\eeqa
with shorthand notations $\rydd \equiv {d\ryd\over d\t}$ and $\delta\rydd \equiv {d\delta \ryd\over d\t}$.
The first and second line in (\ref{acceleration_anomalies}) contain all the terms which are not modulated
by the annual motion of Earth. In particular, they contain the secular contribution to the anomalous Doppler
acceleration which was calculated in \cite{JR05cqg,JR06cqg}.
Note the relative signs between terms in first and second lines of (\ref{acceleration_anomalies}),
which correct an error made in \cite{JR05cqg,JR06cqg}.
The third line contains modulated terms depending locally on anomalies of the probe trajectory and 
shown below to give negligeable contributions to Pioneer-like anomalies.
The fourth line includes terms which depend on the non local anomaly $\db\ri$. 
These important terms (see below) were ignored in \cite{JR05cqg,JR06cqg}.
They vary with the Earth motion around the Sun and determine annual modulations of the anomaly. 

The Pioneer anomaly has been recorded on deep space probes with high excentricity orbits and, 
therefore, nearly radial motions. For the sake of simplicity, we neglect from now on all the terms 
proportional to angular velocities or angular accelerations.
As proper time relativistic corrections depend on the probe velocity squared $\ryd^2$, 
it is also possible to use the simplification $d\s \simeq \c d\t$.
In this context, one deduces from (\ref{velocity_normalization}) the radial acceleration 
read as the sum of standard and anomalous contributions
\beqa
\label{probe_acceleration}
\stand{\rydd} &\simeq& {\c^2\over2} {\partial_\r\ln\stand{\g_{00}}\over\stand{\g_{\r\r}}}
-{\stand{\ryd}^2\over2}\partial_\r\ln\stand{\g_{00}\g_{\r\r}}\nonumber\\
\delta\rydd &\simeq& {\c^2\over2} {\db(\partial_\r\g_{00})\over\stand{\g_{00}\g_{\r\r}}}
- \stand{\rydd }{\db(\g_{00}\g_{\r\r})\over\stand{\g_{00}\g_{\r\r}}}\nonumber\\
&&-{\stand{\ryd}^2\over2}{\db(\partial_\r(\g_{00}\g_{\r\r}))\over\stand{\g_{00}\g_{\r\r}}}
-{\partial_\r\stand{\g_{00}\g_{\r\r}}\over\stand{\g_{00}\g_{\r\r}}}\stand{\ryd}\delta\ryd
\eeqa
We have disregarded the effect of planets as gravity sources,
which is justified once again by the fact that we focus our attention on the anomalies 
(this effect has to be taken into account in the data analysis process \cite{Anderson02}). 
Equation (\ref{velocity_normalization}) is integrated to obtain the anomaly expressed on 
the position or velocity of the probe
\beqa
\delta\ryd &\simeq& {\c^2\db\g_{00}/2 -\c^2\stand{\er}\delta \er \over\stand{\g_{00}\g_{\r\r}}\stand{\ryd}}
-{\db (\g_{00}\g_{\r\r})\over2\stand{\g_{00}\g_{\r\r}}}\stand{\ryd} \nonumber\\
\delta\ry &\simeq& \stand{\ryd}\left\lbrace \delta\tz+\int_\rx^\ry {\c^2\delta\g_{00} - 2\c^2\stand{\er}\delta\er  -
\stand{\rd}^2 \delta(\g_{00}\g_{\r\r})\over2\stand{\rd}^3\stand{\g_{00}\g_{\r\r}}} d\r
 \right\rbrace
\eeqa
These relations are just particular cases of the more general results obtained in the end of the
preceding section. They have been written in terms of a difference $\delta\tz$ of epoch, that is
also the time of passage at the initial radius $\r\tin$. For convenience, this initial radius 
has been pushed back to $\rx$.

The Doppler acceleration anomaly (\ref{acceleration_anomalies})
may then be expressed in terms of the radial velocity $\stand{\ryd}$ and 
acceleration $\stand{\rydd}$
\beqa
\label{Pioneer_anomaly}
\db\ad &=& {\c^2\stand{\tdr}\over2}{\db(\partial_\r\g_{00})\over\stand{\g_{00}\g_{\r\r}}}\nonumber\\
&-&\stand{\rydd}\left\lbrace{\stand{\tdr}\db(\g_{00}\g_{\r\r})\over\stand{\g_{00}\g_{\r\r}}}
+{\delta(\g_{00}\g_{\r\r})\over2\stand{\g_{00}}^2\stand{\tdr}}
-{\stand{\g_{\r\r}}\delta\g_{00}\over\stand{\g_{00}}^2\stand{\tdr}}\right\rbrace\nonumber\\
&-&\stand{\ryd}^2\left\lbrace {\stand{\tdr}\over2}\left({\db(\partial_\r(\g_{00}\g_{\r\r}))
\over\stand{\g_{00}\g_{\r\r}}}-{\partial_\r\stand{\g_{00}\g_{\r\r}}
\over\stand{\g_{00}\g_{\r\r}}^2}\db(\g_{00}\g_{\r\r})\right)\right.\nonumber\\
&&\quad +\left.\partial_\r\left({\delta(\g_{00}\g_{\r\r})\over2\stand{\g_{00}^2\tdr}}\right)
+{1\over2\stand{\ryd}}{d\stand{\tdr}\over d\t}{\db(\g_{00}\g_{\r\r})\over\stand{\g_{00}\g_{\r\r}}}
-\partial_\r\left({\stand{\g_{\r\r}}\delta\g_{00}\over\stand{\g_{00}^2\tdr}}\right)\right\rbrace\nonumber\\
&+&\left\lbrace {1\over\stand{\ryd}}{d\stand{\tdr}\over d\t}
-\stand{\tdr}{\partial_\r\stand{\g_{00}\g_{\r\r}}\over\stand{\g_{00}\g_{\r\r}}}\right\rbrace
{\c^2\db\g_{00} -2\c^2\stand{\er}\delta \er\over2\stand{\g_{00}}\stand{\g_{\r\r}}}\nonumber\\
&-& {\stand{\ri}\stand{\dot{\ri}}\stand{\ryd}\over\stand{\ry}^2\stand{\g_{00}}^2\stand{\tdr}^3}
\left\lbrace{\delta(\g_{00}\g_{\r\r})\over2}-\stand{\g_{\r\r}}\delta\g_{00}\right\rbrace
\nonumber\\
&+& {d\over d\t}\left\lbrace\stand{\ryd}\stand{\partial_\r\tdr}\delta\ry 
+\stand{\ri}\delta\dang + {\stand{\rid} \over \stand{\rip}} \db\ri\right\rbrace
\eeqa
We now evaluate this expression by using orders of magnitude known for the different quantities.
We first know that the metric components are close to 1 for $\g_{00}$, to -1
for $\g_{\r\r}$, the differences being of the order of the Newton potential $\Gauss/\r$.
The latter has a value $\sim 10^{-8}$ on Earth orbit
and values 20 to 70 times smaller at the distances explored by Pioneer probes. 
The square of velocity divided by light velocity has the same value $\sim 10^{-8}$
for Earth, due to the virial theorem, and it is roughly 6 times smaller for Pioneer
probes (velocity $\sim$ 12 km/s $\sim$ 0.4 times that of Earth).
As we study deep space probes at large heliocentric distances $\ry \gg \rx$, we 
also use the fact that the parameter $\stand{\ri}$ is at least 20 times smaller
than $\stand{\ry}$, so that terms scaling as $\stand{\ri}^2/\stand{\ry}^2$ 
are at least 400 times smaller than unity. 
For the same reason, terms proportional to angular velocity anomaly 
$\delta\dang$ are found to have a negligible effect.  
After these remarks, expression (\ref{Pioneer_anomaly}) is simplified to
the following dominant contributions
\beqa
\label{simplified_Pioneer_anomaly}
&&\db\ad \simeq \db\ad _\sec + \db\ad_\ann\\ 
&&\db\ad _\sec \simeq -{\c^2\over2} \partial_\r(\delta\g_{00})
+\stand{\rydd}\left\lbrace{\delta(\g_{00}\g_{\r\r})\over2} -\delta\g_{00}\right\rbrace
-{\c^2\over2}\partial_\r^2\stand{\g_{00}}\delta\ry \nonumber\\
&&\db\ad _\ann \simeq {d\over d\t}\left\lbrace\stand{\dang}\db\ri\right\rbrace \nonumber
\eeqa
The term $\db\ad _\sec$ contains secular contributions proportional to metric anomalies 
in the first and second sectors as well as to the range ambiguity
\beqa
\label{range_ambiguity}
&&\delta\ry \simeq \stand{\ryd}\left\lbrace \delta\tz-\int_\rx^\ry {\c^2\delta\g_{00} 
- 2\c^2\stand{\er}\delta\er  - \stand{\rd}^2 \delta(\g_{00}\g_{\r\r}) \over2\stand{\rd}^3}d\r 
\right\rbrace
\eeqa
The last term $\db\ad _\ann$ is a modulated contribution which is
proportional to the anomaly of the impact parameter
\beqa
&&\db\ri\simeq -\stand{\rip}\stand{\ri} \left\lbrace \int_\rx^{\ry}
\left({\delta(\g_{00}\g_{\r\r})\over2} +\delta\g_{00}\right) {d\r\over\r^2}
+ {\delta\ry\over\stand{\ry}^2}\right\rbrace 
\eeqa
At this point, it is worth emphasizing the differences between the expression
(\ref{simplified_Pioneer_anomaly}) obtained for the anomalous acceleration $\db\ad$ 
and the result previously obtained in \cite{JR05mpl,JR05cqg}. 

The secular anomaly arising from the second sector
replaces the result obtained in preliminary calculations
\cite{JR05mpl,JR05cqg} which was spoiled by a calculation error.  
The previous result was linear in the gravity fields and
proportional to the kinetic energy of the probe. 
The new expression appears at the second order in gravity fields, since
it is proportional on one hand to the anomalous potential $\delta(\g_{00}\g_{\r\r})$ 
at the position of the probe and on another hand to the standard 
probe acceleration $\stand{\rydd}$, that is also the gradient of the standard Newton
potential. It however remains of first order with respect to the 
metric anomaly, and this property has been used in the derivation.
Note that the change of form of this term has no consequence on the comparison of 
the anomalies recorded on Pioneer 10 and 11 since the two probes had nearly equal velocities 
and anomalous accelerations \cite{Anderson02}.
But it will affect several conclusions to be drawn in the next section. 

The previous results \cite{JR05mpl,JR05cqg} were also preliminary for the following reasons.
First, the secular anomaly is corrected by a term proportional to the range ambiguity, 
because the position of the probe is not known directly
(no range measurement capabilities on Pioneer probes),
and this important fact was not discussed previously. 
The range ambiguity (\ref{range_ambiguity}) contains contributions proportional to
anomalies as well as trajectory mismodeling, \textit{i.e.} modifications
of the standard acceleration observable due to changes $\delta\tz$ and $\delta\er$ 
of the constants of motion.
Then, the range ambiguity also affects the evaluation of the annually modulated 
anomaly, probably the most striking new feature of the expression 
(\ref{simplified_Pioneer_anomaly}).
Expectations for the annual and secular anomalies are thus correlated,
a property which will turn out to be of uttermost importance in the next section.
Finally, the comparison of these expectations may open a road to a genuine
test of the `post-Einsteinian' phenomenological framework, as discussed below.

\section{Discussion of orders of magnitude}

Obviously, the interest of this new road depends in a critical manner on the orders
of magnitude of the various terms, to be discussed in the present section.
This discussion is heavily dependent on the stringent constraints put on possible 
metric anomalies by the gravity tests already performed in the solar system.
In order to write down the relevant arguments, we introduce potentials $\PN$ and $\PP$ 
in the two sectors as in \cite{JR05cqg,JR06cqg}
\beqa
\g_{00}&\simeq& 1+2\PN \quad,\quad
-\g_{00}\g_{\r\r}\simeq 1+2\PP 
\eeqa
We also use the simplest form of the standard impact parameter $\stand{\ri}$ 
which is modulated by Earth rotation 
\beqa
\label{modulated_form}
&&\stand{\ri}\simeq -\rx\sin\left(\Ox(\t-\t\tconj)\right)
\eeqa
The time $\t\tconj$ corresponds to a conjonction (closest approach) of Earth and deep space probe.
The simple expression (\ref{modulated_form}) is sufficient for the purpose of the present section.

In this context, the secular and modulated anomalies in (\ref{simplified_Pioneer_anomaly}) 
are reduced to 
\beqa
\label{oversimplified_Pioneer_anomaly}
&&\db\ad _\sec \simeq -\c^2 \partial_\r\delta\PN (\ry) 
+\aE {\rx^2\over\ry^2} \left\lbrace { 2\delta\ry \over\ry} + \delta\PP (\ry) + 2\delta\PN (\ry) \right\rbrace
\nonumber\\
&&\db\ad_\ann \simeq -\annfactor\cos\left(2\Ox(\t-\t\tconj)\right) 
-{d\annfactor\over2\Ox d\t} \sin\left(2\Ox(\t-\t\tconj)\right) 
\nonumber\\
&&\annfactor \equiv \aE{\rx\over\ry} \left\lbrace
{\delta\ry\over\ry} - \ry \int_\rx^{\ry} (\delta\PP-2\delta\PN) {d\r\over\r^2} 
\right\rbrace\nonumber\\
&&\aE\equiv {\c^2 \kappa\over\rx^2} = {\GN \M\over\rx^2} = \Ox^2\rx \simeq 6\times10^{-3}\mathrm{m\,s}^{-2}
\eeqa
These expressions of the secular and modulated anomalies can be considered as the key predictions
of the post-Einsteinian framework presented in this paper. 

The secular anomaly $\db\ad_\sec$ contains a first term describing the effect of a 
Newton law modification and a second one gathering the contributions of the range 
ambiguity and of the potentials.
The annual anomaly $\db\ad_\ann$ is determined by an amplitude $\annfactor$ 
which contains contributions of the range ambiguity and of the two anomalous potentials. 
This amplitude is not annually modulated but suffers a secular change during the 
probe's journey. Considering that $\annfactor$ varies slowly over a year, the 
modulation in (\ref{oversimplified_Pioneer_anomaly}) is essentially 
at twice the orbital frequency. The observed behaviour \cite{Anderson02} has a richer
structure but its discussion must take into account the following remarks. First the
constants of motion, and therefore $\annfactor$, are modified at maneuvers
which are occuring twice a year on the average. Then, the mismodeling contributions 
on angles $\angy$ and $\angly$, not studied in detail here,
have periods $\Ox$ and $2\Ox$, as soon as a non null inclination angle $\iota$
is accounted for. These remarks qualitatively explain why the observed annual modulation 
may be more complicated than the simple expression (\ref{oversimplified_Pioneer_anomaly}).

The maneuvers are not frequent for Pioneer 10/11 probes, which is one of the main causes
of their excellent navigational accuracy \cite{Anderson02}. 
They interrupt the free geodesic segments by changing the values of the 
constants of motion. This essentially amounts to a change of the 
local velocity with no change of the position if the maneuver can be modeled as nearly 
instantaneous. The detailed description of the maneuvers is one of the most delicate parts
of the data analysis process and it certainly goes much farther than the purpose of the
present paper. In the following paragraphs, we focus our attention on the main features
which can be qualitatively expected rather than on quantitative results. To this aim,
we can go along with the simplified expression (\ref{oversimplified_Pioneer_anomaly}).

The hierarchy of magnitudes appearing in (\ref{oversimplified_Pioneer_anomaly}),
$\aE \rx^2/\ry^2$ in $\db\ad_\sec$ and $\aE \rx/\ry$ in $\db\ad_\ann$,
plays a central role in this discussion.
This is the reason why we have used in (\ref{oversimplified_Pioneer_anomaly})
the acceleration $\aE$ of Earth on its orbit as the natural scale for measuring 
anomalous Doppler accelerations, except for the first term which is
the direct effect of a Newton law modification. 
We briefly discuss this term now, before embarking on a more complete analysis 
of the terms induced by the range ambiguity and the second potential 
and putting into evidence the correlation between the secular and modulated anomalies.

Should the Pioneer anomaly be explained by an anomaly in the first sector, 
a linear dependence of the potential $\delta\PN$ would be needed to reproduce the fact that
the anomaly has a roughly constant value (\ref{Pio_anom}) over a large range of 
heliocentric distances $\rP$
\beqa
\label{hypot_N}
\c^2 \partial_\r\delta\PN \simeq\aP
\quad,\quad 20\,\mathrm{AU}\le\rP\le70\,\mathrm{AU} 
\eeqa
The simplest way to modelize the anomaly would thus correspond to a potential varying 
linearly with $\r$ and vanishing at Earth orbit to fit the convention (\ref{null_at_Earth})
\beqa
\label{model_N}
\delta\PN\simeq {\r-\rx\over\lP}
\eeqa
We have introduced a length $\lP$ characteristic of the Pioneer anomaly
\beqa
\label{length_P}
\lP^{-1}\equiv{\aP\over\c^2} \simeq 0.8\times10^{-26}\,\mathrm{m}^{-1}
\simeq 1.2\times10^{-15}\,\mathrm{AU}^{-1} 
\eeqa

Should this model effectively describe the metric in the vicinity of Earth and Mars,
its effects could not have escaped detection in the very accurate tests 
performed with martian probes such as Viking \cite{Reasenberg79}. 
Numbers sheding light on this point are given in \cite{Anderson02} 
(see also \cite{JR04} where similar conclusions are obtained 
for a modification of the Newton potential having the form of a Yukawa potential). 
The effect of the perturbation (\ref{model_N}) on planets would produce a change of their 
orbital radius. The order of magnitude (\ref{length_P}) would lead to range variations of
$\sim$50km and $\sim$100km respectively at smallest and largest distances.
Meanwhile, the Viking data constrain these measurements to agree with standard expectations
at a level of $\sim$100m and $\sim$150m respectively.
These numbers are different enough to eliminate the simple model (\ref{model_N}) with the 
coefficient $\lP$ chosen to fit the Pioneer anomaly.
Note that the effect of the Shapiro time delay in the range evaluation, which should in 
principle have been taken into account in the discussion, has here a negligible influence
\cite{Moffat06}.

As already discussed, these results do not prove that the Pioneer anomaly 
cannot be reproduced in a metric theory. 
First, there is the possibility that the linear dependence needed to reproduce (\ref{hypot_N}) 
at distances explored by Pioneer probes is cut at the orbital radii of planets on which the 
strongest constraints are obtained \cite{Brownstein06}. 
As it was already discussed,  
it then remains to decide whether or not the ephemeris of the outer planets are accurate enough 
to forbid the presence of the linear dependence (\ref{hypot_N}) in 
the range of distances explored by the Pioneer probes \cite{Iorio06,Tangen06}.
This point remains to be settled \cite{Brownstein06}. 
In any case, there is another possibility, namely that the Pioneer anomaly
is induced by the second anomalous potential 
$\delta\PP$ rather than the first one $\delta\PN$.
We now consider these terms which are still here even if there is no anomaly at all 
in the first sector ($\delta\PN=0$).
More thorough studies will have to be performed later on to study
the correlated effects of anomalies in the two sectors. 

We now focus our attention on the secular anomaly $\db\ad _\sec $ and 
the annual amplitude $\annfactor$ which are determined in 
(\ref{oversimplified_Pioneer_anomaly}) by the range ambiguity and the second potential
\beqa
\label{oversimplified_second_sector}
&&\db\ad _\sec \simeq 
\aE {\rx^2\over\ry^2} \left\lbrace { 2\delta\ry \over\ry} + \delta\PP (\ry) \right\rbrace
\,,\,
\annfactor \simeq \aE{\rx\over\ry} \left\lbrace
{\delta\ry\over\ry} - \ry \int_\rx^{\ry} {\delta\PP \over\r^2} d\r 
\right\rbrace
\eeqa
With the same assumption $\delta\PN=0$, the range ambiguity is given by
\beqa
\label{oversimplified_ambiguity}
&&\delta\ry \simeq \stand{\ryd} \left\lbrace\delta\tz 
+ \int_\rx^\ry {\delta\enr \over\stand{\rd}^3}d\r 
- \int_\rx^\ry {\delta\PP \over\stand{\rd}}d\r \right\rbrace
\eeqa
We have introduced the non relativistic reduced energy $\enr$ 
which allows us to express the standard velocity $\stand{\rd}$
as a function of distance, using (\ref{velocity_normalization}), 
\beqa
\label{vinfty}
&&\enr\equiv{\c^2\left(\er^2-1\right)\over2} \quad,\quad 
\stand{\rd} = \sqrt{2 \left(\enr + {\c^2\Gauss\over\r} \right)}
\eeqa

The secular anomaly and annual amplitude appear in (\ref{oversimplified_second_sector}) 
as different linear superpositions of the reduced range ambiguity $\delta\ry/\ry$ and
of the second potential $\delta\PP$. The term in front of the annual amplitude turns out
to be larger than the term in front of the secular anomaly by a factor $\ry/\rx$.
This could appear to be contradictory with the fact that the annual anomaly is only a 
fraction of the secular one \cite{Anderson02} in the data, but this is not the case.
As a matter of fact, the data analysis process described in \cite{Anderson02}
is based on the \textit{a priori} assumption that there
is no annually modulated anomaly in the physical signal of interest. 
As explained above, this assumption is valid for anomalies induced by $\delta\PN$
but not for anomalies induced by $\delta\PP$.
In the context of this assumption, the choice of the best trajectory fitting the data
tends to produce a null or quasi null value for the annual anomaly.
This corresponds to a choice of the trajectory leading to a perfect or nearly perfect 
compensation of the two contributions to $\annfactor$ in (\ref{oversimplified_second_sector}),
in conformity with the assumed absence of annual anomalies.

In the context of the present paper where annual anomaly of the Doppler acceleration is a 
natural expectation, the observations reported in \cite{Anderson02} 
have a different significance.
They mean that the unknown epoch $\delta\tz$ characterizing the 
motion of the probe has been fixed so that the two contributions to $\annfactor$ 
compensate each other at some well chosen radius $\ry$
\beqa
\label{first_order_compensation}
&&\left|\annfactor\right| \ll \aE{\rx\over\ry} \quad,\quad
{\delta\ry\over\ry} \simeq \ry\int_\rx^{\ry} {\delta\PP \over\r^2} d\r 
\eeqa
As a consequence of already presented arguments, this compensation has to be effective
within a fraction of the order of or even smaller than $\rx/\ry$. 
Now, this compensation cannot remain perfect over a long period of time.
The first reason for that is the different dependence on $\ry$ of the two terms to be
compensated by each other. The second reason is due to the maneuvers which change the constants
of motion and thus affect the compensation.
A precise estimation of the annual anomaly thus requires a complete solution of the
motion including a detailed description of the maneuvers.
As this task is outside the scope of the present paper, we will not be able to conclude
whether or not the annual modulations reported in \cite{Anderson02} are effectively 
accounted for in a quantitative manner by the effect of the second potential.

Despite this deficiency, the description just given of the annual anomaly nevertheless 
leads to a quantitative estimation of the secular anomaly.
Equation (\ref{first_order_compensation}) indeed fixes the otherwise
unknown range ambiguity $\delta\ry/\ry$, so that $\db\ad_\sec$ can be rewritten
\beqa
&&\db\ad _\sec \simeq 
\aE \rx^2 \left\lbrace {2\over\ry} \int_\rx^{\ry} \chP(\r) d\r  + \chP(\ry)  
\right\rbrace \,,\quad  \delta\PP(\r) \equiv \chP(\r) \r^2
\eeqa
A roughly constant anomaly is produced when $\chP$ is constant, 
\textit{i.e.} when $\delta\PP(\r)$ is quadratic in $\r$,
in the range of Pioneer distances.
Identifying the expression $\db\ad_\sec$ to the observed Pioneer anomaly $\aP$
fixes the value of the constant 
\beqa
\label{hypot_P}
\db\ad_\sec \simeq -\aP \quad\rightarrow\quad \chP\simeq -{1\over3\Gauss\lP} 
\simeq -4\times10^{-8}\,\mathrm{AU}^{-2} 
\eeqa
This value is 3 times smaller from what would have been obtained
without accounting for the range ambiguity. Note that
$\chP$ can take different values outside the range 20-70 AU of Pioneer distances
and that it is not even forced to be exactly constant in this range.
In fact, we know that $\chP$ has to vanish at Earth radius in order to obey 
the convention (\ref{null_at_Earth}). We also show in the next paragraph that 
it may have to be smaller than (\ref{hypot_P}) between Earth and Mars in order 
to be compatible with planetary observations.

To this aim, we consider a simple model with the potential obtained as the
sum of linear and quadratic terms vanishing at Earth orbit to fit (\ref{null_at_Earth})
\beqn
\label{model_P}
\delta\PP \simeq -{(\r-\rx)^2+\mP(\r-\rx)\over3\Gauss\lP} 
\eeqn
The quadratic coefficient has been fixed according to (\ref{hypot_P}).
The further characteristic length $\mP$ has been introduced to represent 
the radial derivative of the metric anomaly at Earth orbit.
This linear term has to be small enough to be dominated by the quadratic one 
at distances explored by Pioneer probes ($\mP\ll\rP$).
Now the metric perturbation (\ref{model_P}) has also an influence
on the already discussed range measurements on martian probes.
As a consequence of the Shapiro effect, a range variation is found with a value
of the order of $\sim$700m, in conflict with Viking data.
This conflict may be cured by cutting off the simple dependence (\ref{model_P}) 
at the orbital radii of Earth and Mars.
It is easily checked out that this does not affect significantly the predictions 
made for the Pioneer probes which are at much larger distances.

\section{Concluding remarks}

As stated in the introduction, the present paper follows up publications
\cite{JR05mpl,JR05cqg,JR06cqg}
devoted to the study of post-Einsteinian metric extensions of GR.
A more complete theory of radar ranging and Doppler tracking has been given 
in terms of the time delay function, allowing us to discuss the annual anomaly 
besides the secular one. A mistake made in previous publications in the 
evaluation of the secular anomaly has been corrected and important
new results have been obtained. 
In particular, the annual anomaly has been found to be correlated with the secular 
anomaly through terms arising from ambiguities in the position of the probe.
This correlation, in principle available in the data, has to be 
scrutinized in order to extract reliable information from Pioneer observations.

The following qualitative statements summarize the results of
the present paper. As the secular anomaly, the annual anomaly is a natural 
consequence of the presence of a second potential. 
This has to be contrasted with the first potential which 
does not produce a significant effect along the links \cite{Moffat06}. 
Then, the annual anomaly produced by propagation along the up- and down-links 
can be compensated near an arbitrary point by an appropriate choice of the 
trajectory of the probe. In fact this compensation is an output of any best fit
procedure based on the a priori assumption that there is no annual anomaly.
However, the compensation cannot remain exact when the probe moves as the
two compensating terms have different dependences on the 
heliocentric distance. It follows that the annual anomaly reappears, either 
after some free evolution or after the next maneuver.
This qualitative behaviour is reminiscent of the observations of
annual anomalies which were reported in \cite{Anderson02}.
This situation certainly pleads for pushing this study and comparing the 
theoretical expectations with Pioneer data. 
It is only after a quantitative comparison, taking into account all the details
known to be important for data analysis \cite{Anderson02}, that it will be possible to
decide whether or not the post-Einsteinian phenomenological framework does fit
the observations. 

It is clear after these remarks that some of the conclusions of our previous papers 
have to be amended: the secular anomaly turns out to be proportional to the
standard acceleration and to the second potential.
The corrected expression is quadratic, and no longer linear, in the gravity fields,
with one contribution standard and the second one anomalous.
Identifying the expectation with the observed Pioneer anomaly now points to a 
second potential with a quadratic dependence on the radius. 
This corresponds to a constant curvature with an unexpectedly large value
in the outer solar system (eq. (\ref{hypot_P}) of the present paper).
This quadratic dependence may have to be cut off at distances exceeding
the size of the solar system as well as in the inner solar system
(in order to pass Shapiro tests on martian probes).
Note that the discussion of the preceeding section was mainly 
focused on the change of the Shapiro delay, due to the anomaly on $\PP$.
The modification of the orbital radii, which could in principle play a role, 
has been ignored because it was expected to have a negligible influence. 
The evaluation of correlated effects of anomalies in the two sectors will
be necessary in order to be able to reach definitive conclusions. 

These conclusions constitute motivations for new experiments in the solar system.
Clearly, experiments with ranging capabilities will offer qualitatively better 
perspectives than Pioneer observations which were performed without such capabilities.
Missions going to the borders of the solar system \cite{Pioneer05} will either prove 
or disprove the existence of the anomaly at such long distances.
Comparison with the theoretical expectation of the present paper will give an 
answer to the question whether such an anomaly may have a metric origin,
with the metric possibly departing from the GR prediction.
This idea could also be tested on a shorter time scale by adding 
specially designed instruments on planetary probes going to Mars, Jupiter, or Saturn,
the reduction of the explored heliocentric distance being compensated by 
a potentially large improvement of the measurement accuracy.

\def\etal{\textit{et al }}
\def\ibid{\textit{ibidem }}
\def\eprint#1{{\it Preprint} #1}


\begin{thebibliography}{99}
\bibitem{Schw}  An equivalent representation of the same spacetime is given in 
 \cite{JR06cqg} in terms of Schwartzschild coordinates
\end{thebibliography}
\end{document}